\documentclass[aps,prl,twocolumn,showpacs,groupedaddress,letterpaper]{revtex4}
\usepackage{graphicx}
\usepackage{amssymb}
\usepackage{amsmath}
\newlength{\figwidth}
\begin{document}
\preprint{\today}
\title{NRG Study of an Inversion-Symmetric Interacting Model: \\
Universal Aspects of its Quantum Conductance}

\author{Axel Freyn}
\altaffiliation[Present address:]{ Institut N\'eel, 25 avenue des 
Martyrs, BP 166, 38042 Grenoble, France.}
\affiliation{Service de Physique de l'\'Etat Condens\'e (CNRS URA 2464), 
IRAMIS/SPEC, CEA Saclay, 91191 Gif-sur-Yvette, France}

\author{Jean-Louis Pichard}
\affiliation{Service de Physique de l'\'Etat Condens\'e (CNRS URA 2464), 
IRAMIS/SPEC, CEA Saclay, 91191 Gif-sur-Yvette, France}

\begin{abstract} 
We consider scattering of spinless fermions by an inversion-symmetric 
interacting model characterized by three parameters (interaction $U$, 
internal hopping $t_d$ and coupling $t_c$). Mapping this spinless model 
onto an Anderson model with Zeeman field, we use the numerical 
renormalization group for studying the particle-hole symmetric case. 
We show that the zero temperature limit is characterized by a line of 
free-fermion fixed points and a scale $\tau(U,t_c)$ of $t_d$ for which 
there is perfect transmission. The quantum conductance and the 
low energy excitations of the model are given by universal functions 
of $t_d/\tau$ if $t_d < \Gamma$ and of $t_d/t_c^2$ if $t_d > \Gamma$, 
$\Gamma = t_c^2$ being the level width of the scatterer. 
This universal regime becomes non-perturbative when $U$ exceeds $\Gamma$.
\end{abstract}

\maketitle

 In quantum transport theory, the conductance $G$ of a nanosystem  
inside which the electrons do not interact is given by 
$g=G/(e^2/h)= |t_{\it ns}|^2$ when the temperature $T \to 0$, 
$|t_{\it ns}|^2$ being the probability for an electron at the Fermi energy 
$E_F$ to be transmitted through the nanosystem. This Landauer-Buttiker 
formula can be extended to an interacting nanosystem, if it behaves 
as a non-interacting nanosystem with renormalized parameters. We study 
such a renormalization using the numerical renormalization 
group (NRG) algorithm \cite{hewson,krishna-murthy2} and an 
inversion-symmetric interacting model (ISIM) which describes the scattering 
of spin-polarized electrons (spinless fermions) by an interacting region 
characterized by an internal hopping term $t_d$, a coupling term $t_c$ 
and an interaction strength $U$. This model was used \cite{asada,fkp}
for studying the effect of an external scatterer upon the 
effective transmission of an interacting region, assuming the Hartree-Fock 
(HF) approximation. We revisit ISIM with the NRG algorithm for investigating 
non-perturbative regimes where other methods (NRG or DMRG 
algorithms) than the HF approach become necessary.  
 
 Quantum impurity models \cite{hewson}, as the Anderson model which 
describes a level with Hubbard interaction $U$ coupled to a 3d bath of free 
electrons, were introduced to study the resistance 
minimum observed in metals with magnetic impurities. The Kondo problem 
refers to the failure of perturbative techniques to describe this  
minimum. The solution of these models by the NRG algorithm, 
a non-perturbative technique \cite{hewson,krishna-murthy2} introduced by 
Wilson, is at the origin of the discovery of universal behaviors which 
can emerge from many-body effects. The observation \cite{goldhaber-gordon} 
of the Kondo effect in semiconductor quantum dots has opened a second 
era for quantum impurity models, now used for modeling mesoscopic 
objects (single \cite{silvestrov} or double \cite{borda} quantum dot 
systems) inside which electrons interact, in contact with baths of free 
electrons (large conducting non interacting leads). 

Though the Kondo effect is induced by magnetic moments, it is also 
at the origin of spinless models, such as the interacting resonant 
level model \cite{mehta} (IRLM) which describes a resonant level 
($V_d d^{\dagger}d$) coupled to two baths of spinless electrons via 
tunneling junctions and an interaction $U$ between the level and 
the baths. IRLM, which is often used for studying nonequilibrium 
transport \cite{mehta,boulat}, is related to the Kondo model, the 
charge states $n_d=0,1$ playing the role of spin states. 
Both ISIM and IRLM are inversion symmetric and can exhibit orbital 
Kondo effects. However, the Zeeman field acting on the impurity is 
played by the hopping term $t_d$ for ISIM, and by the site energy 
$V_d$ for IRLM. Therefore, ISIM does not transmit the electrons 
without field, while IRLM does. The two-particle states have been 
given for ISIM \cite{dhar}.

 For the particle-hole symmetric case \cite{krishna-murthy2}, the 
Anderson model maps onto the Kondo Hamiltonian if $U>\pi \Gamma$, 
$\Gamma$ being the impurity-level width. In that case, there is a 
non-perturbative regime where the temperature 
dependence of physical observables such as the impurity susceptibility 
is given by universal functions of $T/T_K$, $T_K$ being the Kondo 
temperature. If $U < \pi \Gamma$, the impurity susceptibility can be 
obtained by perturbation theory. Mapping ISIM onto an Anderson model 
with a Zeeman field $t_d$, and assuming that the role of $t_d$ should 
qualitatively resemble that of a finite temperature, we expect the following 
scenario for the ISIM conductance $g$ of the particle-hole symmetric 
case: If $U > \pi \Gamma \propto t_c^2$, we expect a non-perturbative 
regime where $g$ should be given by a universal function of $t_d/\tau$ 
independently of the values of $U$ and $t_c$, with a scale $\tau(t_c,U)$ of 
$t_d$ playing the role of a Kondo temperature $T_K$. If $U < \pi \Gamma$, the 
HF theory should correctly give $g$. This scenario will be more or 
less confirmed by extensive NRG calculations.
\begin{figure}
\centerline{
\includegraphics[width=0.8\columnwidth]{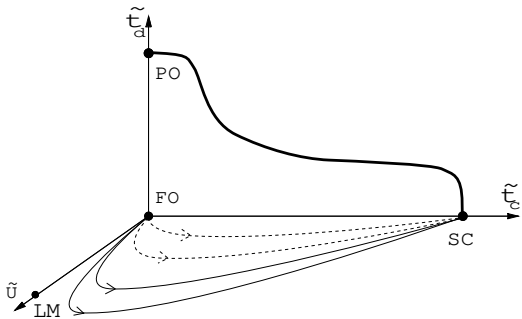}
}
\caption
{Line of free fermion fixed points (${\tilde U}=0$, thick solid line) 
characterizing ISIM when $T \to 0$ as $t_d$ increases from $t_d=0$ 
(SC fixed point) towards $t_d \to \infty$ (PO fixed point). The FO, LM 
and SC fixed points and the RG trajectories \cite{hewson} followed by 
ISIM as $T$ decreases for $t_d=0$ are indicated in the plane 
$\widetilde{t_d}=0$, for $\pi \Gamma > U$ (dashed) and 
$\pi \Gamma <U $ (solid).
}  
\label{fig1} 
\end{figure}
\begin{figure*}
\centerline{
\includegraphics[width=\textwidth]{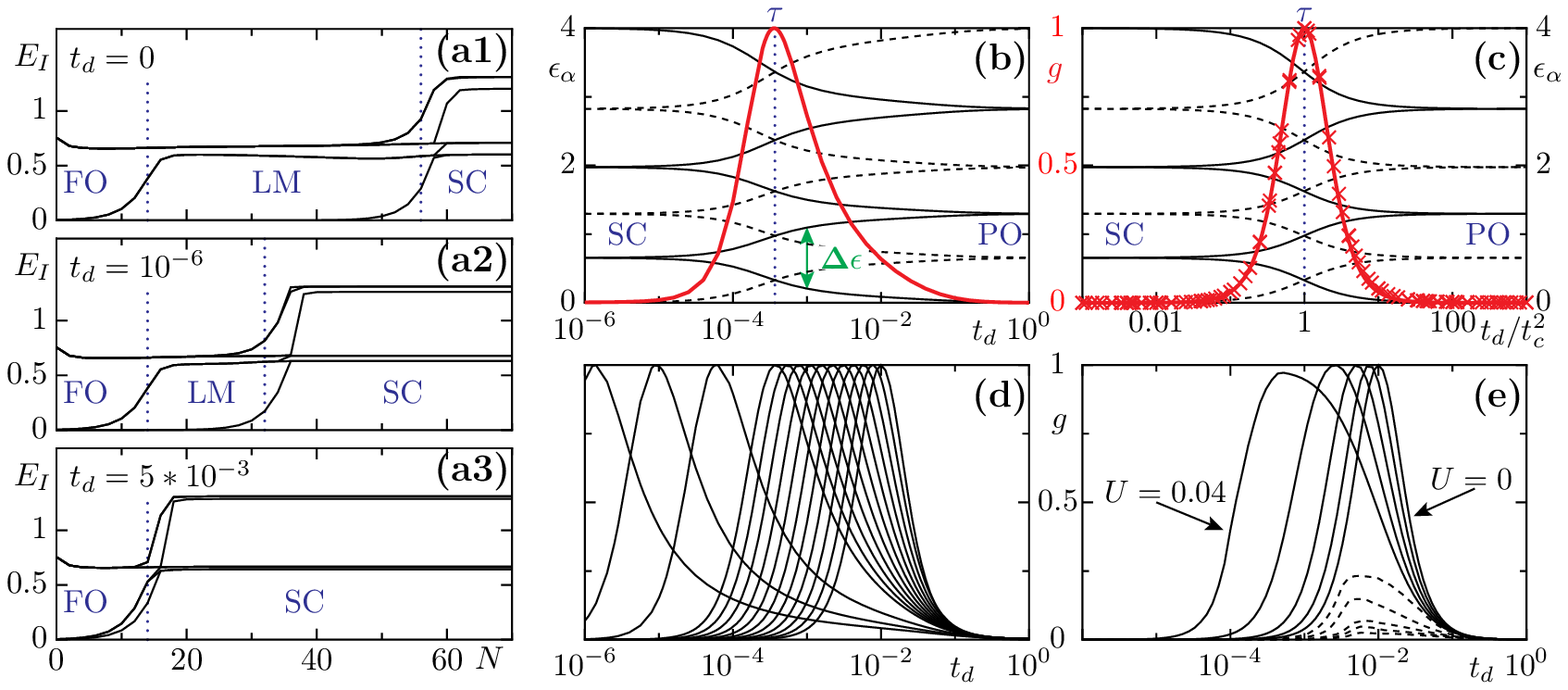}
}
\caption{(Color online) Fig.~\ref{fig2}(a): Many body excitations $E_{I}$ 
as a function of $N$ (even values) for  $U=0.005$ and $t_c=0.01$. For $t_d=0$ 
(Fig.~\ref{fig2}(a1)), one can see the 3 successive plateaus 
(FO, LM and SC fixed points) of the Anderson model~\cite{krishna-murthy2}. 
As $t_d$ increases (Fig.~\ref{fig2}(a2) and Fig.~\ref{fig2}(a3)), the LM 
plateau shrinks and disappears when $t_d \approx U$. Fig.~\ref{fig2}(b): 
One body excitations $\epsilon_{\alpha} (t_d)$ (extracted from the 
$E_{I}(N \to \infty, t_d)$) for $U=0.1$ and $t_c=0.1$ (left scale). 
The solid (dashed) line corresponds to NRG chains of even (odd) length $N$. 
Conductance $g(t_d)$ extracted from $\Delta \epsilon(t_d)$ using 
Eq.~\eqref{transmission} (thick red curve, right scale). 
For $t_d=\tau$, the $\epsilon_{\alpha}$ are 
independent of the parity of $N$ and $g=1$. Fig.~\ref{fig2}(c): For $U=0$, 
$\epsilon_{\alpha}(t_d/t_c^2)$ and $g(t_d/t_c^2)$ extracted from the NRG 
spectra ($\times$). $g=\cosh^{-2}(X)$ (red line) with $X=\ln(t_d/t_c^2)$ is 
correctly reproduced. 
Figs.~\ref{fig2}(d),(e): $g(t_d)$ for $t_c=0.1$ and 
many values of $U$, calculated by NRG algorithm (d) and by 
HF theory (e). In Fig.~\ref{fig2}(d), the larger is $U$, the smaller 
is $t_d=\tau$ where $g=1$. The curves correspond respectively to $U=0.25, 
0.2, 0.15$ (3 left peaks) and $U=0.1, 0.09, \ldots, 0.01, 0$ (11 right 
peaks). In Fig.~\ref{fig2}(e), the HF values are accurate for $U=0.02, 0.01, 
0$ (3 right peaks), but become inaccurate when $U \approx 0.04 \approx 
\Gamma$. For $U > \Gamma$, the HF curves (dashed lines) are very different 
of the corresponding NRG curves (Fig.~\ref{fig2}(d)).  
\label{fig2}
}
\end{figure*}

{\it ISIM Hamiltonian:} $H=H_{\it ns}+H_{\it l}+H_{\it c}$. The 
Hamiltonian of the interacting region (the nanosystem) reads:
\begin{equation}
\label{eq_ham_S}
H_{ns} = - t_d \left( c_0^\dagger c_1^{\vphantom{\dagger}} + c_1^\dagger
c_0^{\vphantom{\dagger}} \right ) 
+ V_G \left( n_0 + n_1 \right ) + U n_0 n_1\,.
\end{equation}
$c_x^\dagger$ and $c_x^{\vphantom{\dagger}}$ are spinless fermion operators 
at site $x$ and $n_x=c_x^\dagger c_x^{\vphantom{\dagger}}$. The leads are 
described by an Hamiltonian $H_{\it l}= - t_h 
\sideset{}{'}{\sum}_{x=-\infty}^{\infty} ( c_x^\dagger 
c_{x+1}^{\vphantom{\dagger}} + H.c.)$, where $\sideset{}{'}{\sum}$ means 
that $x=-1,0,1$ are omitted from the summation. The coupling Hamiltonian 
$H_{\it c}=- t_c ( c_{-1}^\dagger c_0^{\vphantom{\dagger}} + c_1^\dagger 
c_2^{\vphantom{\dagger}} + H.c. )$.

{\it Mapping onto an Anderson model with Zeeman field:} Because of 
inversion symmetry, one can map ISIM onto a semi-infinite 1d 
lattice where the fermions have a pseudo-spin and 
the double site nanosystem becomes a single site with Hubbard repulsion 
$U$ at the end point of the semi-infinite lattice. 
$a_{\mathrm e/\mathrm o,x}^{\dagger} = (c_{-x+1}^{\dagger} \pm c_x^{\dagger})
/{\sqrt 2}$ creating a spinless fermion in an even/odd 
($\mathrm e/\mathrm o$) combination of the orbitals at the 
sites $x$ and $-x+1$ of the infinite lattice, (or a fermion with 
pseudo-spin $\sigma=\mathrm e/ \mathrm  o$ in a semi-infinite lattice), 
one gets $H_{ns} = (V_G - t_d) n_{\mathrm e} + 
(V_G + t_d) n_{\mathrm o} + U n_{\mathrm e} n_{\mathrm o}$, 
where $n_{\sigma}=a_{\sigma,1}^{\dagger} 
a_{\sigma,1}^{\vphantom{\dagger}}$ and where the pseudo-spin ``$\mathrm  e$'' 
(``$\mathrm o$'') is parallel (anti-parallel) to the 
``Zeeman field'' $t_d$. In terms of the operators 
$d^{\dagger}_{k,\sigma}=\sqrt{2/\pi} \sum_{x=2}^{\infty} 
\sin( k (x-1)) a^{\dagger}_{\sigma,x}$ creating a spinless fermion of 
pseudo-spin $\sigma$ and momentum $k$ in the semi-infinite lattice, 
$H_{\it l} = 
\sum_{k,\sigma} \epsilon_k n_{k,\sigma}$ and 
$H_{\it c} = \sum_{k,\sigma} V (k) ( a_{\sigma,1}^\dagger 
d_{k,\sigma}^{\vphantom\dagger}+ H. c. )$, where the $k$-dependent 
hybridization $V(k)=-t_c \sqrt{2/\pi} \sin k$ yields an impurity level 
width $\Gamma=t_c^2$, $n_{k,\sigma}= 
d_{k,\sigma}^\dagger d_{k,\sigma}^{\vphantom\dagger}$ 
and $\epsilon_k=-2t_h \cos k$. ISIM is almost the Anderson model, 
except that the impurity has a Zeeman field $t_d$ and is coupled 
to a semi-infinite 1d bath of free electrons. When $t_d \to 0$, ISIM 
exhibits an orbital Kondo effect if the equivalent Anderson model can 
be reduced to a Kondo model.

{\it NRG procedure:} ISIM can be studied using Wilson's procedure 
\cite{hewson,krishna-murthy2} developed for the Anderson model 
after minor changes. First, we assume $V(k) \approx V(k_F=\pi/2)$ 
and, taking $\Lambda = 2$, we divide the conduction band (logarithmic 
discretization) of the electron bath  
into sub-bands characterized by an index $n$ and an energy width 
$d_n=\Lambda^{-n}(1-\Lambda^{-1})$. Within 
each sub-band, we introduce a complete set of orthonormal functions 
$\psi_{np} (\epsilon)$, 
and expand the lead operators in this basis. Dropping the terms with 
$p \neq 0$ and using a Gram-Schmidt procedure, the original 1d leads   
give rise to another semi-infinite chain with nearest neighbor hopping 
terms, each site being labelled by the same index $n$ as the energy 
sub-band from which it comes, and representing a conduction electron 
excitation at a length scale $\Lambda^{n/2} k_F^{-1}$ centered 
on the impurity. In this transformed 1d model, the successive sites 
are coupled by hopping terms $t_{n,n+1} \propto \Lambda^{-n/2}$ 
which vanish as $n \to \infty$. The impurity and the $N-1$ first sites 
form a NRG chain of length $N$ and of Hamiltonian $H_N$. This length can be 
interpreted \cite{krishna-murthy2} as a logarithmic temperature scale. 
The NRG chain coupled to the impurity is iteratively diagonalized and 
rescaled, the spectrum being truncated to the $N_s$ first 
states at each iteration. The behavior of ISIM as $T$ decreases can be 
obtained from the 
spectrum of $H_N$ as $N$ increases, the bandwidth of $H_N$ being suitably 
rescaled at each step. A fixed point of the RG flow corresponds to an 
interval of successive even (or odd) values of $N$ where the rescaled 
many-body excitations $E_{I}(N)$ do not vary. If it is a free-fermion 
fixed point, $E_{I}=\sum_{\alpha} \epsilon_{\alpha}$, the 
$\epsilon_{\alpha}$ being one-body excitations, and the interacting system 
behaves as a non-interacting system ($\tilde U=0$) with renormalized 
parameters $\widetilde{ t_d}$ and $\widetilde{t_c}$ near the fixed point. 
Moreover, if one has free fermions when $T \to 0$, $g$ can be extracted 
from the NRG spectrum. 

{\it Symmetric case:} Using this NRG procedure, ISIM can be studied 
as a function of $T$ for arbitrary values of its bare parameters. 
Hereafter, we take $t_h=1$, $E_F=0$ and $V_G=-U/2$. This choice makes 
ISIM invariant under particle-hole symmetry, with a uniform density 
$\left(\langle n_x\rangle=1/2\right)$ and 3 effective parameters 
$( {\tilde U}, \widetilde{t_c}, \widetilde{t_d})$.

{\it Suppression of the LM fixed point as $t_d$ increases:}
When $t_d=0$, ISIM is an Anderson model which has the RG flow 
sketched in Fig.~\ref{fig1} for the particle-hole symmetric case. 
At low  values of $N$ (high values of $T$), ISIM is located in 
the vicinity of the unstable free orbital (FO) fixed point. As 
$N$ increases ($T$ decreases), ISIM flows towards the stable strong 
coupling (SC) fixed point. If $\pi t_c^2 < U$, the flow can visit an  
intermediate unstable fixed point: the local moment (LM) fixed point 
before reaching the SC fixed point. In that case, ISIM is identical to a 
Kondo model characterized by a temperature $T_K$ and by universal 
functions of the ratio $T/T_K$. If $\pi t_c^2 > U$, the flow goes 
directly from the FO fixed point towards the SC fixed point, and there 
is no orbital Kondo effect for $t_d \to 0$. In Fig.~\ref{fig2}(a),  
the first many-body excitations $E_{I}$ of ISIM are given for increasing 
even values of $N$ for $t_d=0$. Since $\pi t_c^2 < U$, one gets 3 plateaus 
corresponding to the 3 expected fixed points. Inside the plateaus, the 
spectra are free-fermions spectra which are described in 
Ref.~\cite{krishna-murthy2}. However, between the plateaus, there are 
no free-fermion spectra and $E_{I}\neq \sum_{\alpha} \epsilon_{\alpha}$. 
As $t_d$ increases (Fig.~\ref{fig2}(a)), the LM plateau decreases and 
vanishes when $t_d \approx U$.

{\it Evolution of the SC fixed point as $t_d$ increases:}
In the limit  $N \to \infty$ ($T \to 0$), let us study the ${E_{I}}$ 
as a function of $t_d$. For $t_d=0$, one has the SC 
limit~\cite{krishna-murthy2} where the impurity is strongly coupled to 
the second site (the conduction-electron state at the impurity site) of 
the NRG chain. The impurity and this site form a system which can 
be reduced to its ground state (a singlet), the $N-2$ other sites carrying 
free fermions excitations $\epsilon_{\alpha}$ which are independent of that 
system. In the presence of a Zeeman field $t_d \neq 0$, the free-fermion rule 
$E_{I}(t_d)=\sum_{\alpha} \epsilon_{\alpha} (t_d)$ remains valid (see 
Fig.~\ref{fig2}(b)) and the $T \to 0$ limit of ISIM is given by a continuum 
line of free-fermion fixed points where $\tilde U=0$, as sketched in 
Fig.~\ref{fig1}. When the pseudo-spin degeneracy is broken, the first 
(second) one-body excitation $\epsilon_1$ ($\epsilon_2$) carry respectively 
an even (odd) pseudo-spin if $N$ is even. This is the inverse if $N$ is odd, 
$\epsilon_1$ ($\epsilon_2$) carrying respectively an odd (even) pseudo-spin. 
For $t_d \to \infty$, the impurity occupation numbers $n_{\mathrm e}=1$ and 
$n_{\mathrm o}=0$, and the $N-1$ other sites of the NRG chain are 
independent of the impurity. We call this fixed point ``Polarized  Orbital'' 
(PO), since it coincides with the FO fixed point of the Anderson model, 
except that the spin of the free orbital is fully polarized in our case. 
Since for $N \to \infty$ and $t_d \to 0$ (SC fixed point), the free part of 
the NRG chain has $N-2$ sites, while it has $N-1$ sites for $t_d \to \infty$ 
(PO fixed point), there is a permutation of the ${\epsilon_{\alpha}(t_d)}$ 
as $t_d$ increases: as shown in Figs.~\ref{fig2}(b) and (c),  
the ${\epsilon_{\alpha}(t_d\to 0)}$ for $N$ even become the 
${\epsilon_{\alpha}(t_d \to \infty)}$ for $N$ odd and vice-versa. 

{\it Characteristic energy scale $\tau$:}
We define the characteristic energy scale $\tau (t_c,U)$ of ISIM as 
the value of $t_d$ for which the $\epsilon_{\alpha} (t_d)$ are 
independent of the parity of $N$ when $N \to \infty$. Because of 
particle-hole symmetry, the nanosystem (the impurity of the NRG chain) 
is always occupied by one electron. Binding one electron of the leads 
with this electron reduces the energy when $t_d<\tau$, while it increases 
the energy when $t_d > \tau$. For $t_d=\tau$, it is indifferent to bind 
or not an electron of the lead with the one of the nanosystem, making 
ISIM perfectly transparent. This gives the proof that, for every  values 
of $U$ and $t_c$, there is always a value $\tau$ of $t_d$ for which 
$g=1$. The argument is reminiscent to that giving the condition for having 
a perfectly transparent quantum dot in the Coulomb blockade regime: $t_d$ 
in our case, the gate voltage in the other case, have to be adjusted 
to values for which it costs the same energy to put an extra electron 
outside or inside the dot. 

{\it Extraction of the conductance $g$ from the NRG spectra:} 
If $\delta_e$ ($\delta_o$) are the even (odd) scattering phase shifts at 
$E_F$, 
\begin{equation}
g(t_d)=\sin^2(\delta_e - \delta_o) =\sin^2
\left(\pi \frac{\Delta \epsilon(t_d)}
{\Delta\epsilon(t_d \to \infty)}\right),
\label{transmission} 
\end{equation}
where $\Delta \epsilon=\epsilon_2-\epsilon_1$ is the energy gap 
between the two first excitations of a NRG chain of even length 
$N \to \infty$ (see Fig.~\ref{fig2}(b)). When $U=0$, this relation is a 
consequence of Friedel sum rule, which can be written for each 
pseudo-spin channel separately. In that case, $g=\cosh^{-2} (X)$ where 
$X=\ln(t_d/t_c^2)$ and $\tau=t_c^2$. The $\epsilon_{\alpha}(t_d)$ 
given by the NRG algorithm for $U=0$ are shown in Fig.~\ref{fig2}(c) 
with the corresponding values of $g$ obtained from  Eq.~\eqref{transmission}, 
showing that this procedure gives correctly $g$ when $U=0$. It has been 
shown~\cite{borda,hofstetter,oguri} that Eq.~\eqref{transmission} 
can also be used when $U \neq 0$, if there are free fermions when 
$T \to 0$. 

{\it Non-perturbative regime ($U > \Gamma/A)$}: In HF theory, $t_d$ 
takes~\cite{asada} a value $v=t_d+U \langle c_0^\dagger 
c_1^{\vphantom{\dagger}} (v,t_c) \rangle$ and $g=1$ if $v=t_c^2$. 
This gives for the scale $\tau$ a HF value $\tau_{HF}=t_c^2 - A U$ where 
$A=\langle c_0^\dagger c_1^{\vphantom{\dagger}} (v=t_c^2,t_c) \rangle$ 
depends weakly on $t_c$, $A=1/\pi$ ($1/4$) for $t_c=1$ ($0$). When $U 
\to t_c^2/A$, $\tau_{HF} \to 0$, showing that HF theory cannot be used 
above an interaction threshold which is almost the threshold 
$\pi \Gamma$ giving the onset of the non-perturbative regime for the 
Anderson model. 
This breakdown of HF theory for $U \approx \Gamma/A$ can be seen if one 
compares Fig.~\ref{fig2}(d) (NRG results) and Fig.~\ref{fig2}(e) (HF results).
\begin{figure}
\centerline{
\includegraphics[width=\columnwidth]{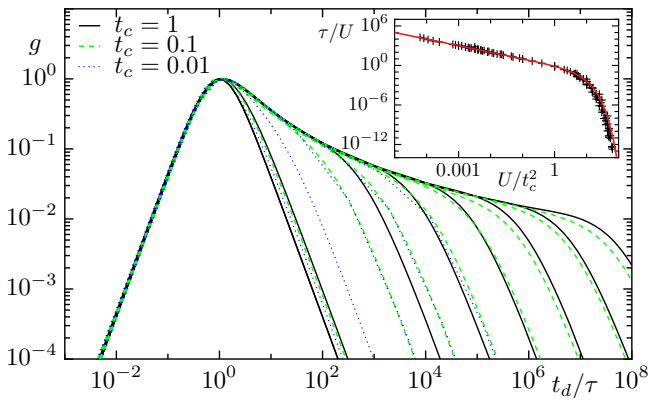}
}
\caption{(Color online) Conductance $g$ as a function of $t_d/\tau$ for 3 values 
of $t_c$ and many values $0 \leq U \leq 35$. Inset: $\tau(U,t_c)/U$ 
as a function of $U/t_c^2$ (+) and fit $\tau= t_c^2 \exp-(U/(\pi t_c^2))$ 
(solid red line).
}  
\label{fig3} 
\end{figure}

{\it Universality:} The conductance $g$ extracted from the NRG 
spectra for $t_c=0.01, 0.1$ and $1$ and $0 \leq U \leq 35$ is 
given as a function of $t_d/\tau$ in Fig.~\ref{fig3}. One can see 
3 successive regimes. When $t_d < \tau$, there is a single curve which 
is independent of $U$ and $t_c$ and which corresponds to $g=\cosh^{-2} 
(X)$ with $X=\ln(t_d/\tau)$, and not $\ln(t_d/t_c^2)$ as for $U=0$. When 
$t_d > \tau$, another universal curve independent of $t_c$ and 
$U$ describes the data as a function of $t_d/\tau$ as far as $t_d$ 
does not exceed $\Gamma$. Indeed, the same data plotted 
as a function of $t_d$ show that $g$ becomes independent of $U$ when 
$t_d > \Gamma$. In this third regime (parallel lines which can be seen 
in Fig.~\ref{fig3} for large values of $t_d/\tau$) $g=\cosh^{-2} (X)$ with 
$X=\ln(t_d/t_c^2)$ as if $U=0$. 

{\it Roles of $T$ and $t_d$:} 
We have assumed analogies between the effect of $T$ in the Anderson model, 
the effect of a Zeeman field at $T=0$ in the Anderson model, and eventually 
the effect of $t_d$ at $T=0$ in ISIM. This was based on the 
idea that the singlet state of the SC limit could be broken either if the 
temperature $T$ or the Zeeman energy $t_d$ exceeds $T_K$. Let us summarize 
the interest and the limit of these analogies. Increasing $T$ in 
the Anderson model (or in ISIM with $t_d=0$), one gets 3 regimes, each of 
them being characterized by a single fixed point (Fig.~\ref{fig2}(a)). There 
are no free fermions for temperatures $T \approx T_K$ (SC--LM crossover) 
and $T \approx \Gamma$ (LM--FO crossover). In contrast, increasing  
$t_d$ in ISIM at $T=0$, one has always free fermions (Fig.~\ref{fig2}(b)), 
and not only around 3 fixed points. However, there are 3 regimes in ISIM 
as $t_d$ increases, as in the Anderson model as $T$ increases, 
delimited by 2 energy scales $\tau$ and $\Gamma$. The behavior of 
$\tau \approx t_c^2 \exp-(U/(\pi t_c^2))$ (inset of Fig.~\ref{fig3}) 
resembles that of $T_K \approx t_c \sqrt{\pi U/2} \exp -(U/(\pi t_c^2))$ 
(in ISIM units), 
while the second scale is given by $\Gamma$ in the 2 models. Eventually, 
we point out the similarity between the universality discussed in this 
letter for $g$ and that which characterizes~\cite{kaul} also at $T=0$ 
the behavior of the singlet-triplet gap for a magnetic impurity confined 
in a box of mean level spacing $\Delta$, as a function $T_K/\Delta$.

We thank Denis Ullmo for very useful discussions and the ``Triangle de 
la Physique'' for financial support.  


\begin{thebibliography}{15}
\expandafter\ifx\csname natexlab\endcsname\relax\def\natexlab#1{#1}\fi
\expandafter\ifx\csname bibnamefont\endcsname\relax
  \def\bibnamefont#1{#1}\fi
\expandafter\ifx\csname bibfnamefont\endcsname\relax
  \def\bibfnamefont#1{#1}\fi
\expandafter\ifx\csname citenamefont\endcsname\relax
  \def\citenamefont#1{#1}\fi
\expandafter\ifx\csname url\endcsname\relax
  \def\url#1{\texttt{#1}}\fi
\expandafter\ifx\csname urlprefix\endcsname\relax\def\urlprefix{URL }\fi
\providecommand{\bibinfo}[2]{#2}
\providecommand{\eprint}[2][]{\url{#2}}

\bibitem[{\citenamefont{Hewson}(1993)}]{hewson}
\bibinfo{author}{\bibfnamefont{A.~C.} \bibnamefont{Hewson}},
  \emph{\bibinfo{title}{The Kondo Problem To Heavy Fermions}}
  (\bibinfo{publisher}{Cambridge University Press}, \bibinfo{year}{1993}).

\bibitem[{\citenamefont{Krishna-murthy
  et~al.}(1980)\citenamefont{Krishna-murthy, Wilkins, and
  Wilson}}]{krishna-murthy2}
\bibinfo{author}{\bibfnamefont{H.~R.} \bibnamefont{Krishna-murthy}},
  \bibinfo{author}{\bibfnamefont{J.~W.} \bibnamefont{Wilkins}},
  \bibnamefont{and} \bibinfo{author}{\bibfnamefont{K.~G.}
  \bibnamefont{Wilson}}, \bibinfo{journal}{Phys. Rev. B}
  \textbf{\bibinfo{volume}{21}}, \bibinfo{pages}{1003} (\bibinfo{year}{1980}).

\bibitem[{\citenamefont{Asada et~al.}(2006)\citenamefont{Asada, Freyn, and
  Pichard}}]{asada}
\bibinfo{author}{\bibfnamefont{Y.}~\bibnamefont{Asada}},
  \bibinfo{author}{\bibfnamefont{A.}~\bibnamefont{Freyn}}, \bibnamefont{and}
  \bibinfo{author}{\bibfnamefont{J.-L.} \bibnamefont{Pichard}},
  \bibinfo{journal}{Eur. Phys. J. B} \textbf{\bibinfo{volume}{53}},
  \bibinfo{pages}{109} (\bibinfo{year}{2006}).

\bibitem{fkp}
\bibinfo{author}{\bibfnamefont{A.}~\bibnamefont{Freyn}} \bibnamefont{and}
  \bibinfo{author}{\bibfnamefont{J.-L.} \bibnamefont{Pichard}},
  \bibinfo{journal}{Phys. Rev. Lett.} \textbf{\bibinfo{volume}{98}},
  \bibinfo{pages}{186401} (\bibinfo{year}{2007}{\natexlab{a}});
\bibinfo{author}{\bibfnamefont{A.}~\bibnamefont{Freyn}} \bibnamefont{and}
  \bibinfo{author}{\bibfnamefont{J.-L.} \bibnamefont{Pichard}},
  \bibinfo{journal}{Eur. Phys. J. B} \textbf{\bibinfo{volume}{58}},
  \bibinfo{pages}{279} (\bibinfo{year}{2007}{\natexlab{b}});
\bibinfo{author}{\bibfnamefont{A.}~\bibnamefont{Freyn}},
  \bibinfo{author}{\bibfnamefont{I.}~\bibnamefont{Kleftogiannis}},
  \bibnamefont{and} \bibinfo{author}{\bibfnamefont{J.-L.}
  \bibnamefont{Pichard}}, \bibinfo{journal}{Phys. Rev. Lett.}
  \textbf{\bibinfo{volume}{100}}, \bibinfo{pages}{226802}
  (\bibinfo{year}{2008}).

\bibitem[{\citenamefont{Goldhaber-Gordon
  et~al.}(1998)\citenamefont{Goldhaber-Gordon, Shtrikman, Mahalu,
  Abusch-Magder, Meirav, and Kastner}}]{goldhaber-gordon}
\bibinfo{author}{\bibfnamefont{D.}~\bibnamefont{Goldhaber-Gordon}},
  \bibinfo{author}{\bibfnamefont{H.}~\bibnamefont{Shtrikman}},
  \bibinfo{author}{\bibfnamefont{D.}~\bibnamefont{Mahalu}},
  \bibinfo{author}{\bibfnamefont{D.}~\bibnamefont{Abusch-Magder}},
  \bibinfo{author}{\bibfnamefont{U.}~\bibnamefont{Meirav}}, \bibnamefont{and}
  \bibinfo{author}{\bibfnamefont{M.~A.} \bibnamefont{Kastner}},
  \bibinfo{journal}{Nature (London)} \textbf{\bibinfo{volume}{391}},
  \bibinfo{pages}{156} (\bibinfo{year}{1998}).

\bibitem[{\citenamefont{Silvestrov and Imry}(2007)}]{silvestrov}
\bibinfo{author}{\bibfnamefont{P.~G.} \bibnamefont{Silvestrov}}
  \bibnamefont{and} \bibinfo{author}{\bibfnamefont{Y.}~\bibnamefont{Imry}},
  \bibinfo{journal}{Phys. Rev. B} \textbf{\bibinfo{volume}{75}},
  \bibinfo{pages}{115335} (\bibinfo{year}{2007}).

\bibitem[{\citenamefont{Borda et~al.}(2003)\citenamefont{Borda, Zar\'and,
  Hofstetter, Halperin, and von Delft}}]{borda}
\bibinfo{author}{\bibfnamefont{L.}~\bibnamefont{Borda}},
  \bibinfo{author}{\bibfnamefont{G.}~\bibnamefont{Zar\'and}},
  \bibinfo{author}{\bibfnamefont{W.}~\bibnamefont{Hofstetter}},
  \bibinfo{author}{\bibfnamefont{B.~I.} \bibnamefont{Halperin}},
  \bibnamefont{and} \bibinfo{author}{\bibfnamefont{J.}~\bibnamefont{von
  Delft}}, \bibinfo{journal}{Phys. Rev. Lett.} \textbf{\bibinfo{volume}{90}},
  \bibinfo{pages}{026602} (\bibinfo{year}{2003}).

\bibitem[{\citenamefont{Mehta and Andrei}(2006)}]{mehta}
\bibinfo{author}{\bibfnamefont{P.}~\bibnamefont{Mehta}} \bibnamefont{and}
  \bibinfo{author}{\bibfnamefont{N.}~\bibnamefont{Andrei}},
  \bibinfo{journal}{Phys. Rev. Lett.} \textbf{\bibinfo{volume}{96}},
  \bibinfo{pages}{216802} (\bibinfo{year}{2006}).

\bibitem[{\citenamefont{Boulat et~al.}(2008)\citenamefont{Boulat, Saleur, and
  Schmitteckert}}]{boulat}
\bibinfo{author}{\bibfnamefont{E.}~\bibnamefont{Boulat}},
  \bibinfo{author}{\bibfnamefont{H.}~\bibnamefont{Saleur}}, \bibnamefont{and}
  \bibinfo{author}{\bibfnamefont{P.}~\bibnamefont{Schmitteckert}},
  \bibinfo{journal}{Phys. Rev. Lett.} \textbf{\bibinfo{volume}{101}},
  \bibinfo{pages}{140601} (\bibinfo{year}{2008}).

\bibitem[{\citenamefont{Dhar et~al.}(2008)\citenamefont{Dhar, Sen, and
  Roy}}]{dhar}
\bibinfo{author}{\bibfnamefont{A.}~\bibnamefont{Dhar}},
  \bibinfo{author}{\bibfnamefont{D.}~\bibnamefont{Sen}}, \bibnamefont{and}
  \bibinfo{author}{\bibfnamefont{D.}~\bibnamefont{Roy}},
  \bibinfo{journal}{Phys. Rev. Lett.} \textbf{\bibinfo{volume}{101}},
  \bibinfo{pages}{066805} (\bibinfo{year}{2008}).

\bibitem[{\citenamefont{Hofstetter and Zarand}(2004)}]{hofstetter}
\bibinfo{author}{\bibfnamefont{W.}~\bibnamefont{Hofstetter}} \bibnamefont{and}
  \bibinfo{author}{\bibfnamefont{G.}~\bibnamefont{Zarand}},
  \bibinfo{journal}{Phys. Rev. B} \textbf{\bibinfo{volume}{69}},
  \bibinfo{pages}{235301} (\bibinfo{year}{2004}).

\bibitem[{\citenamefont{Oguri and Hewson}(2005)}]{oguri}
\bibinfo{author}{\bibfnamefont{A.}~\bibnamefont{Oguri}} \bibnamefont{and}
  \bibinfo{author}{\bibfnamefont{A.~C.} \bibnamefont{Hewson}},
  \bibinfo{journal}{J. Phys. Soc. Jpn.} \textbf{\bibinfo{volume}{74}},
  \bibinfo{pages}{988} (\bibinfo{year}{2005}).

\bibitem[{\citenamefont{Kaul et~al.}(2006)\citenamefont{Kaul, Zar\'and,
  Chandrasekharan, Ullmo, and Baranger}}]{kaul}
\bibinfo{author}{\bibfnamefont{R.~K.}~\bibnamefont{Kaul}},
  \bibinfo{author}{\bibfnamefont{G.}~\bibnamefont{Zar\'and}},
  \bibinfo{author}{\bibfnamefont{S.}~\bibnamefont{Chandrasekharan}},
  \bibinfo{author}{\bibfnamefont{D.}~\bibnamefont{Ullmo}}, \bibnamefont{and}
  \bibinfo{author}{\bibfnamefont{H.~U.}~\bibnamefont{Baranger}},
  \bibinfo{journal}{Phys. Rev. Lett.} \textbf{\bibinfo{volume}{96}},
  \bibinfo{pages}{176802} (\bibinfo{year}{2006}).

\end{thebibliography}

\end{document}